\documentclass[a4paper,11pt]{article}
\pdfoutput=1 

\usepackage{jinstpub} 
\usepackage{graphicx}
\usepackage{float}
\usepackage{caption}
\usepackage{array}
\usepackage{subfigure}
\usepackage{multirow}
\usepackage{siunitx}
\usepackage{xfrac}
\usepackage{verbatim}
\usepackage{pifont}
\usepackage{arydshln}
\usepackage{paralist}
\usepackage{enumitem}
\usepackage{layout} 
\usepackage{amsmath}
\usepackage{csquotes} 
\usepackage{rotating}
\usepackage{adjustbox}
\usepackage{feynmp}
\DeclareSIUnit\torr{Torr}
\DeclareSIUnit\cee{c}
\DeclareSIUnit\electronvoltnr{eV_r}
\DeclareSIUnit\atomicmassunit{amu}

\title{Measurement of directional range components of nuclear recoil tracks in a fiducialised dark matter detector \boldmath}
\collaboration{The DRIFT Collaboration:}

\author[a]{J.~B.~R.~Battat,}
\author[b]{~E.~J.~Daw,}
\author[1,b,c,d]{A.~C.~Ezeribe,\note{Corresponding author.}}
\author[d]{~J.~-L.~Gauvreau,}
\author[e]{~J.~L.~Harton,}
\author[f]{~R.~Lafler,}
\author[f]{~E.~R.~Lee,}
\author[f]{~D.~Loomba,}
\author[b]{~W.~Lynch,}
\author[g]{~E.~H.~Miller,}
\author[b]{~F.~Mouton,}
\author[c]{~S.~Paling,}
\author[f]{~N.~Phan,}
\author[b]{~M.~Robinson,}
\author[b]{~S.~W.~Sadler,}
\author[b]{~A.~Scarff,}
\author[e]{~F.~G.~Schuckman~II,} 
\author[d]{~D.~P.~Snowden-Ifft}
\author[b]{ and ~N.~J.~C.~Spooner}

\affiliation[a]{Department of Physics, Wellesley College, 106 Central Street, Wellesley, MA 02481, U.S.A.}
\affiliation[b]{Department of Physics and Astronomy, University of Sheffield, S3 7RH, U.K.}
\affiliation[c]{STFC Boulby Underground Science Facility, Boulby Mine, Cleveland, TS13 4UZ, U.K.}
\affiliation[d]{Department of Physics, Occidental College, Los Angeles, CA 90041, U.S.A.}
\affiliation[e]{Department of Physics, Colorado State University, Fort Collins, CO 80523-1875, U.S.A.}
\affiliation[f]{Department of Physics and Astronomy, University of New Mexico, NM 87131, U.S.A.}
\affiliation[g]{Department of Physics, South Dakota School of Mines \& Technology, SD 57701, USA.}
\emailAdd{a.ezeribe@sheffield.ac.uk}

\abstract{We present results from the first measurement of axial range components of fiducialized neutron induced nuclear recoil tracks using the DRIFT directional dark matter detector.  Nuclear recoil events are fiducialized in the DRIFT experiment using temporal charge carrier separations between different species of anions in 30:10:1 \si{\torr} of CS$_2$:CF$_4$:O$_2$ gas mixture. For this measurement, neutron-induced nuclear recoil tracks were generated by exposing the detector to $^{252}$Cf source from different directions. Using these events, the sensitivity of the detector to the expected axial directional signatures were investigated as the neutron source was moved from one detector axis to another. Results obtained from these measurements show clear sensitivity of the DRIFT detector to the axial directional signatures in this fiducialization gas mode.}

\begin{document}
\maketitle
\flushbottom

\section{Introduction}\label{sec:intro}
One of the long-standing tasks in current physics is to unravel the nature of non-baryonic dark matter (DM) \cite{Feng2010} which comprises about 84\si{\percent} \cite{Planck2015} of the mass content of the Universe.  The existence of this non-baryonic DM in the Universe is supported by many observational evidence \cite{Zwicky1933, Ibarra2015,Rubin1970,Bertone2005,Wittman2000,Treu2010,Clowe2006,Markevitch2005,Freese2014,Jedamzik2009,Planck2015,Bertone2010}. Dark matter candidates have been proposed to explain these observed phenomena, but thermal relic particles are the most studied candidates. This is mainly due to their consistent abundance with expectation from DM and their contributions to structure formation in the early Universe \cite{Freese2014,Bertone2010}. From the predictions of the supersymmetric and extra dimension theories, a good particle DM candidate should have a mass range of 10~\si{\giga\electronvolt\per\cee\squared} to 1~\si{\tera\electronvolt\per\cee\squared} \cite{Bergstrom2008,Bergstrom2009}. They are expected to be colourless, long-lived with negligible electromagnetic coupling to standard model particles and can annihilate with their anti-particles. DM candidates with these properties are known as Weakly Interacting Massive Particles (WIMPs).

Detectors have been built to observe WIMP annihilation products without major success \cite{Ahnen2016,Aartsen2013}. Efforts to produce WIMPs in the laboratory using the large hadron collider (LHC) have so far not yielded positive results \cite{Hoh2016,Askew2014}. One of the most promising channels to direct WIMP detection is via the measurement of nuclear recoil tracks from WIMP interactions in sensitive subterranean detectors. The WIMPs are understood to move non-relativistically in our galaxy, and may collide elastically with a target nucleus, creating a low-energy nuclear recoil. Although such interactions are exceedingly rare, there are a number of observable signatures that provide discrimination between WIMPs and backgrounds.  The largest and most robust of these is the $\mathcal{O}(100\%)$ anisotropy in the angular distribution of WIMP-induced nuclear recoils, produced by the motion of the Earth through the WIMP halo~\cite{Spergel1988,Copi2001}. Additionally, the rotation of the Earth modulates this signal at the sidereal rate, creating a smoking-gun signature of dark matter that no known background can mimic.  As direct detection experiements approach the neutrino floor, a high premium is placed on detection strategies that can disentangle a WIMP signature from neutrino backgrounds. Detectors that are sensitive to this directional signature afford an order of magnitude sensitivity increase relative to non-directional experiments~\cite{Grothaus2014,Mayet2016}.

\section{Directional signatures and detection}\label{sec:galacticsign}
A potentially clear and robust signature would be the detection of the direction of WIMP induced nuclear recoil signals \cite{Copi2001,Battat2016}. Such a directional measurement can allow for discrimination of terrestrial, isotropic or solar neutrino backgrounds from WIMP-induced recoils peaked away from Cygnus \cite{OHare2016,Franarin2016,Grothaus2014}.  The DRIFT \cite{Battat2017}, NEWAGE \cite{Nakamura2015}, MIMAC \cite{Riffard2016}, DMTPC \cite{Deaconu2015} and D$^3$ \cite{Vahsen2015} collaborations have developed directional WIMP search time projection chambers. The NEWSdm experiment \cite{Ambrosio2014} has also made progress using nuclear emulsions to measure directions of nuclear recoil tracks. Other methods, for instance columnar recombination in Xenon targets \cite{Nygren2013}, use of polarised $^3$He \cite{Franarin2016}, carbon nanotube \cite{Cavoto2016} and anisotropic crystal scintillator \cite{Cappella2013} targets are being considered. However, there is concern that multiple scattering of nuclear recoil ionization signals in liquid and solid state target detectors can obscure the directional information of tracks \cite{Couturier2017}.  Also, ranges of nuclear recoils are larger in low-pressure gas TPCs, allowing for better track reconstruction, including both the 3D range component R$_3$, and the vector direction (sense) of the recoil.  

\begin{figure*}[b!]
\centering
\includegraphics[clip, trim=0.5cm 8cm 0.5cm 8cm,width=.6\textwidth,height=0.5\textheight,keepaspectratio]{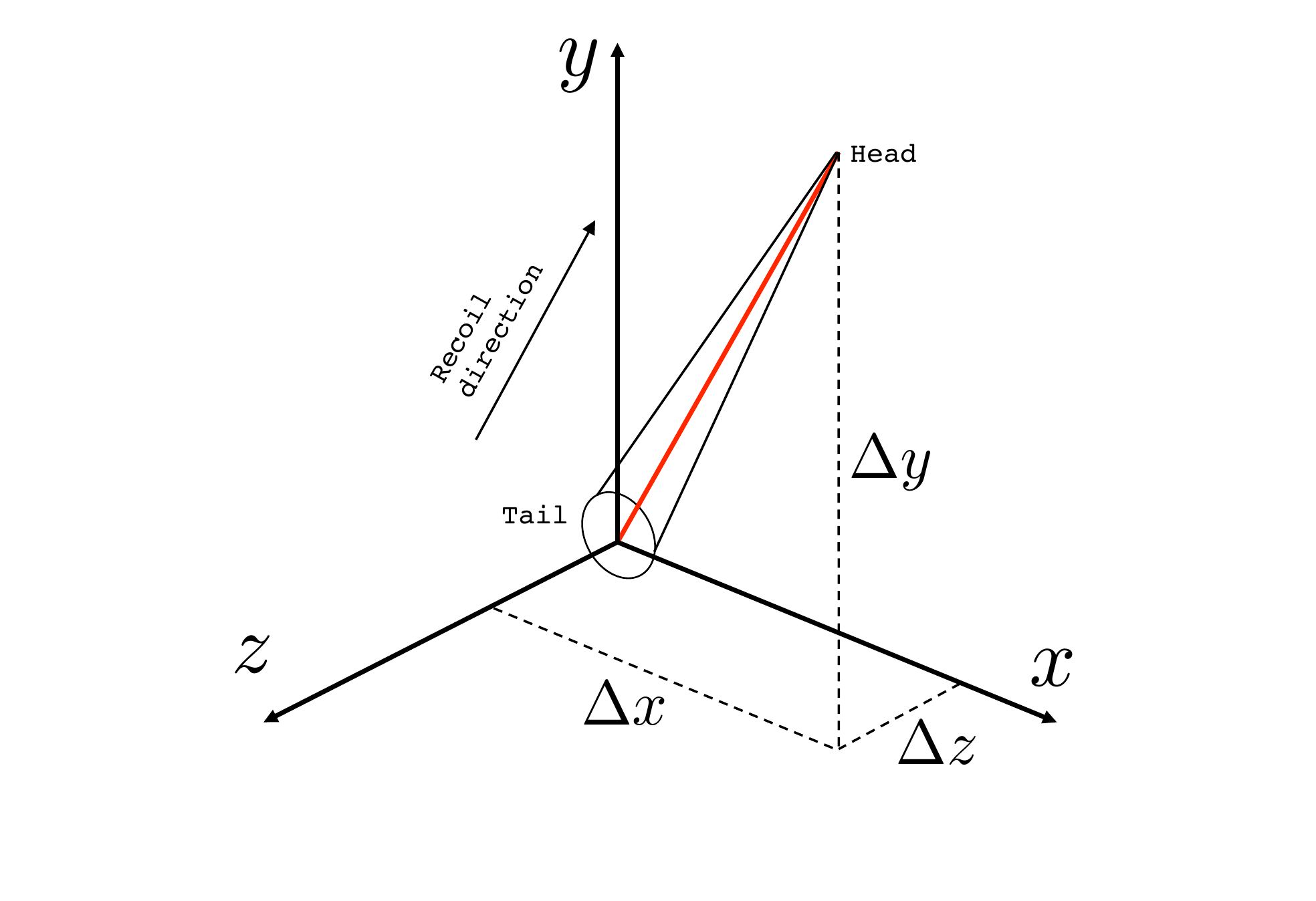}
\caption{Range components of a nuclear track (cone). The cone thickness represents the ionization density along the track. The $\Delta x$, $\Delta y$ and $\Delta z$ axial range components are the projections of the track onto the $x$, $y$ and $z$ axes of the detector, respectively. For clarity, recoil straggling is not included.}
\label{fig:tracksample}
\end{figure*}

\begin{figure*} [b!] 
\centering
\includegraphics[clip, trim=1cm 9cm 1cm 8.5cm,width=0.87\textwidth,height=0.7\textheight,keepaspectratio]{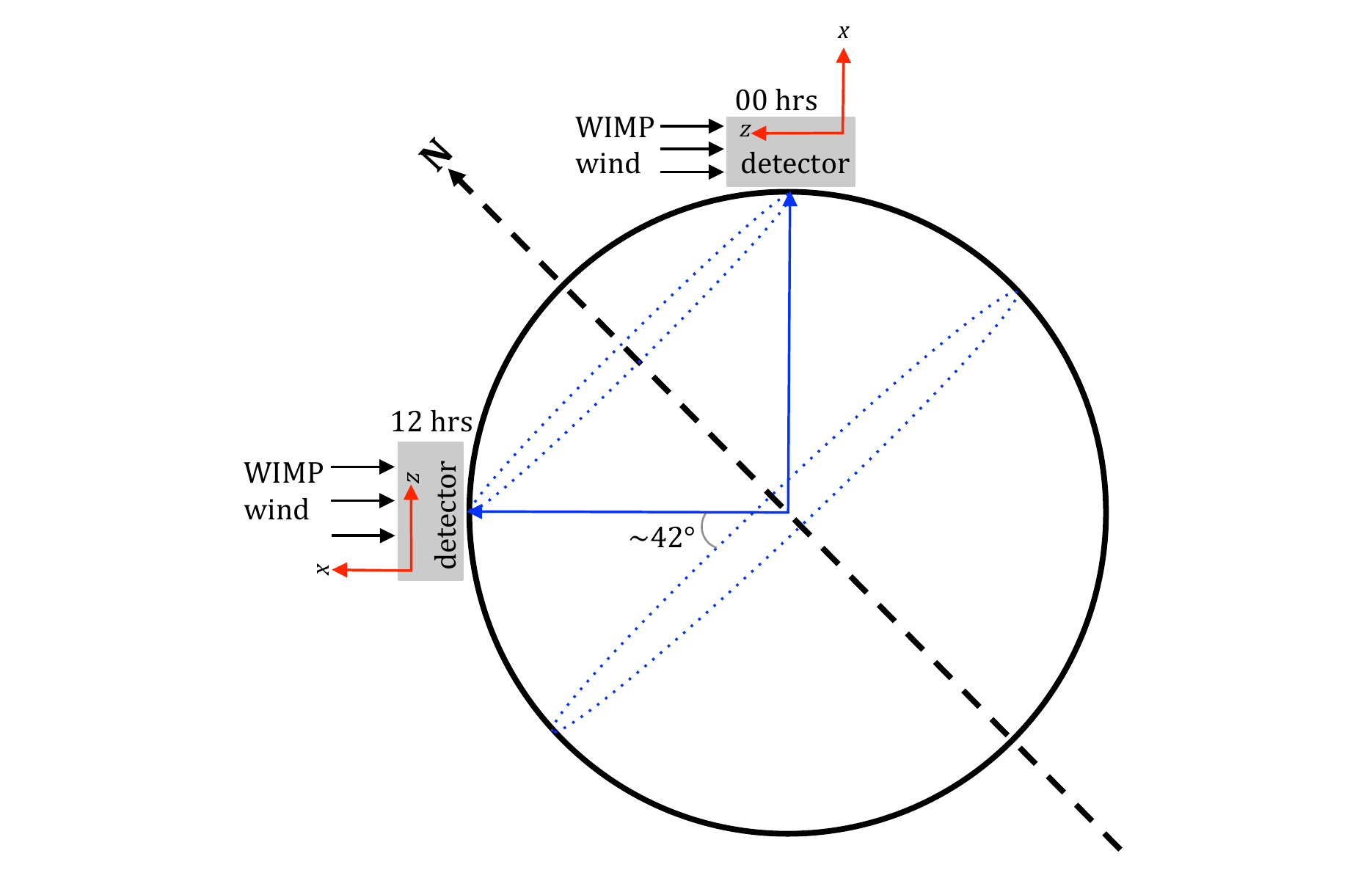}
\caption{Illustration of the expected variation of the WIMP wind direction relative to axes of a detector at 42\si{\degree}N latitude over a sidereal day as the Earth rotates about its axis. The maximum and minimum $\Delta z$ signatures are expected at 00 hrs and 12 hrs, respectively. The red arrows show detector axes, black solid line arrows are the mean WIMP direction, blue single-headed arrows mark the laboratory latitude, dashed solid black line arrow joins the north and south poles of the Earth while dotted blue lines mark latitudes.}
\label{fig:directionalsignal}
\end{figure*}

This vector direction of a nuclear recoil track can be determined from the measurement of charge deposition asymmetry along nuclear recoil tracks, illustrated in Figure \ref{fig:tracksample}. 
This is because the ionization density along a nuclear recoil track decreases from the start (tail) to the end of the track (head). Using this track information, an event vector direction can be deduced \cite{Battat2016,Burgos2009,Spooner2010}.  Another directional galactic WIMP signature can be obtained from measurement of the axial range components $\Delta x$, $\Delta y$, $\Delta z$ of induced nuclear recoil tracks over a sidereal day \cite{Burgos2009a,Snowden-ifft2000} as illustrated in Figure \ref{fig:directionalsignal} for a detector located at 42\si{\degree}N latitude. 
The grey rectangular boxes represent the detector. An ideal detector should be equally sensitive to the $\Delta x$, $\Delta y$, $\Delta z$ range component parameters irrespective of the exposure directions.  However, in detectors with asymmetric axial range component sensitivity like DRIFT, a maximum galactic axial directional WIMP induced signature is expected when the mean WIMP direction is pointing toward the optimal direction. This reduces to a minimum as the WIMP mean direction is in the anti-optimal directions. The optimal direction for the $\Delta x$, $\Delta y$ and $\Delta z$ range components corresponds to the $\widehat{x}$, $\widehat{y}$ and $\widehat{z}$ axes of the DRIFT detector, respectively. The anti-optimal directions for each of the range component parameters are perpendicular to the given optimal direction. Every 12 sidereal hours, the WIMP mean direction changes from optimal to anti-optimal axes of the detector leaving an axial directional signature.  An offset of 13\si{\degree} is expected between the DRIFT detector axes (located at 55\si{\degree}N lattitude) and the mean direction of the WIMP wind from the direction of Cygnus. Even with this offset, it is expected that the DRIFT detector should be sensitive to this directional signature. The capability of the DRIFT detector in measuring these directional signatures has been demonstrated in Refs. \cite{Burgos2009,Burgos2009a} using sulfur recoils at relevant subtended angular spread in 40~\si{\torr} of pure CS$_2$ target gas.  Results from that study suggest that this directional signature will be more useful in the DRIFT-IId design if the $\widehat{z}$ directions of the detector are oriented relative to the Cygnus constellation as illustrated in Figure \ref{fig:directionalsignal}.  Since then, we have added CF$_4$ for spin-dependent WIMP sensitivity \cite{Tovey2000} and O$_2$ for fiducialization \cite{Snowdenifft2014} and now operate with 30:10:1~\si{\torr} of CS$_2$:CF$_4$:O$_2$ mixture. We have shown that the sense recognition capability is preserved with this new gas mixture \cite{Battat2016}. In this work, we seek to show that the range component signature is also preserved.

\section{The DRIFT-IId detector and directed neutron exposures}\label{sec:detectorsandneutronexposures}
\begin{figure*} [b!]  
\centering
\includegraphics[clip, trim=0.5cm 9cm 0.5cm 9cm,width=.8\textwidth,height=0.8\textheight,keepaspectratio]{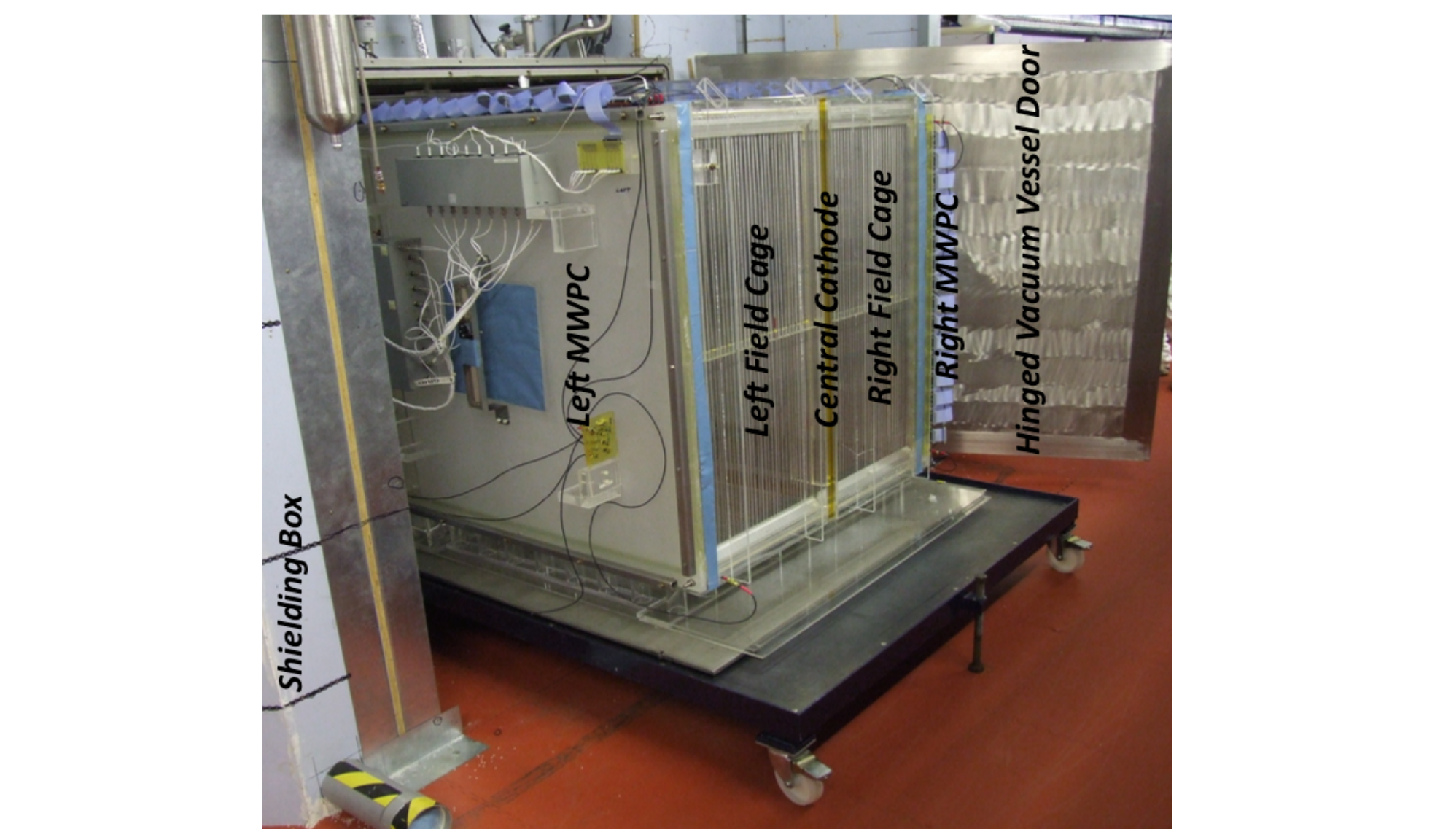}
\caption{The DRIFT-IId detector, partially removed from the stainless steel vacuum vessel and polypropylene pellet shielding (in the grey wooden box). Positions of the central cathode, left and right drift cages, left and right MWPCs, shielding box and the door of the vacuum vessel are labelled. }
\label{fig:driftiiddetector}
\end{figure*}

The DRIFT-IId directional dark matter detector is a 1~\si{\cubic\meter} back-to-back time projection chamber (TPC) which has operated in the STFC underground science facility at the Boulby mine, UK for over a decade \cite{Paling2012, Alner2005}. The DRIFT collaboration has operated different versions of gas based TPCs as directional WIMP search detectors. These include DRIFT-IIa, DRIFT-IIb, DRIFT-IIc  to the present DRIFT-IId detector, shown in Figure \ref{fig:driftiiddetector}.
The two 50-\si{\centi\meter}-long TPC field cages, separated by an aluminized-mylar thin-film central cathode biased at $-$31.9~\si{\kilo \volt}, each support uniform drift fields of 580~\si{\volt \per \centi \meter} \cite{Alner2005,Battat2015,Battat2015b}.  These field cages are housed in a 3.375~\si{\cubic\meter} cubic stainless steel vessel. Each of the two opposite ends of the TPCs is instrumented with a multi-wire proportional chamber (MWPC) \cite{Charpak1979} readout \cite{Alner2005}. Signal avalanche is achieved for each of the MWPCs using a grounded anode wire plane sandwiched between two grid wire planes set at $-$2.884~\si{\kilo\volt} with a 10~\si{\milli\meter} anode-grid distance. Each of the anode and grid wire planes consists of 552 stainless steel wires of 20~\si{\micro\meter} and 100~\si{\micro\meter} diameter, respectively at 2~\si{\milli\meter} pitch. In each of these wire planes, the central 448 wires are grouped down to 8 signal channels such that every 8$^{\text{th}}$ wire is instrumented and read out via one electronics channel. The 16~\si{\milli\meter} distance that is sampled by every 8 contiguous wires in the 8 different signal channels is sufficient to contain \si{\kilo\electronvolt\per\atomicmassunit} nuclear recoil tracks in the DRIFT-IId detector design. Thus, ionzation signals arising from long alpha tracks that traverse more than 8 contiguous wires can cause two or more time-separated charge signals on the same signal channel, allowing for alpha track reconstruction and background discrimination.  The remaining edge wires in every grid plane are grouped down to 1 signal channel and used to veto events entering from outside the detector. The first 22 anode edge wires serve as signal guard wires while the remaining 82 anode edge wires are used as veto for side events. Thus, the anode and grid vetoes fiducialize the detector along its $x$-$y$ dimensions. Signals in each channel are pre-amplified, shaped and filtered using Cremat CR-111, CR-200-4\si{\micro\second} amplifiers and high-pass filters of 110~\si{\micro\second} time constant, respectively.  Two NI PXI-6133 digitizers sample each side of the detector at 1~\si{\mega\hertz} per signal channel. Events on the anode channels are boxcar smoothed over 18~\si{\micro\second}. Consequently, raw events are stored for further analyses on the basis that at least one of the anode smoothed channels must pass a trigger threshold of 15~\si{\milli\volt}.

During operation, the vessel is filled with 30:10:1~\si{\torr} of CS$_2$:CF$_4$:O$_2$ gas mixture. After an interaction in the active volume of the detector, the electronegative CS$_2$ component is used to capture free electrons in the gas.  Drifting ionization track signals as anions minimizes the effect of diffusion to thermal scale \cite{Snowden-Ifft2013}. The high drift field around the MWPC anode wires strips the excess electron from the CS$_{2}^{-}$, leading to proportional avalanche multiplication \cite{Dion2010}. The unpaired spin-$1/2$ proton in the fluorine component of CF$_4$ gas provides the required sensitivity for spin-dependent WIMP-proton coupling searches \cite{Tovey2000}, while the trace of O$_2$ induces the formation of different species of drifting anions needed to reconstruct the position of event vertex from data, known as fiducialization \cite{Snowdenifft2014}. An automated gas handling system is used to maintain high quality of the target gas. This is achieved by mixing the CS$_2$ and CF$_4$+O$_2$ gas components at appropriate pressure with a steady gas flow rate, corresponding to approximately one detector volume change per day.

The gas gain of each of the the two TPC detectors is determined every 6~hrs using 5.9~\si{\kilo\electronvolt} X-ray events from $^{55}$Fe source with remote-controlled shutter mounted behind each of the MWPCs.  For every gas gain calibration operation, data are recorded without any hardware trigger when either the left or right TPC is irradiated.  Data obtained during these calibration runs are analysed for the left and right detector to generate an energy conversion constant for events within the given 6~hrs window \cite{Battat2017}.

\begin{figure*} [t!] 
\centering
\includegraphics[clip, trim=0.4cm 9.5cm 2cm 4.5cm,width=.7\textwidth,height=.6\textheight,keepaspectratio]{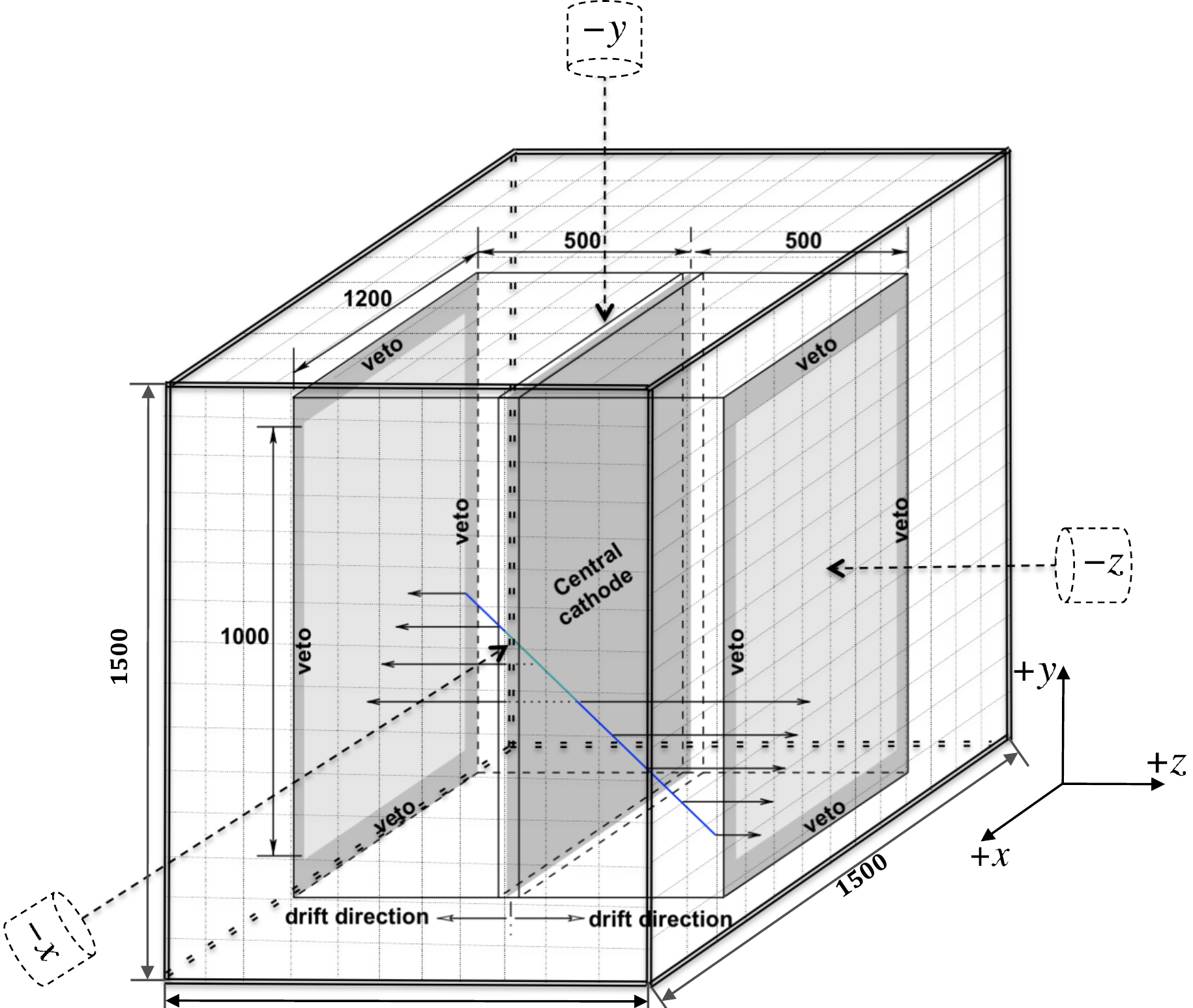}
\caption{Diagram of the DRIFT-IId detector showing the central cathode, readout MWPCs with the veto regions, drift directions, stainless steel vacuum vessel and source positions during the directed neutron exposures. The MWPC anode wires are parallel to $\widehat{y}$, the grid wires are parallel to $\widehat{x}$ and the drift field is parallel to $\widehat{z}$.}
\label{fig:exposuredirections}
\end{figure*}

To investigate the detector sensitivity to the axial directional signature in this operational mode, nuclear recoil tracks generated from the directed $^{252}$Cf \cite{Boulogne1969} neutron exposures reported in Ref. \cite{Battat2016} were used. The isotropic $^{252}$Cf source was positioned such that the mean emitted subtended neutron direction (MND) aligned with the $x$, $y$ and $z$ axes of the detector, as shown in Figure \ref{fig:exposuredirections}. 
The expectation is that the magnitude of the $\Delta z$ range component parameters, for instance, will reach maximum (minimum) when the MND is aligned with $\widehat{z}$ ($\widehat{x}$ or $\widehat{y}$).  As shown in Figure \ref{fig:exposuredirections}, in the $-z$ exposure, the source was placed behind the right MWPC such that the MND pointed toward the central cathode. In the $-x$ ($-y$) exposures, the MND was oriented perpendicular (parallel) to the anode signal wires. The distance between the source and the geometric centre of the central cathode for the $-x$, $-y$ and $-z$ neutron exposures are 1520~\si{\milli\meter},~1520~\si{\milli\meter}~and~620~\si{\milli\meter}, respectively. It is important to point out that due to mechanical reasons, these detector-source distances are smaller relative to our previous work in Ref \cite{Burgos2009a}. As a result, the angular spread and the subtended solid angles of incident neutrons differ in the two measurements. The strength of the axial directional signatures can be improved using neutrons from smaller subtended solid angles. Directed $^{252}$Cf neutron events were used in this measurements since the energy spectrum of neutron-induced nuclear recoil tracks is similar to expectations from WIMPs of relevant energies \cite{Burgos2009}. The polypropylene pellet shielding that surrounds the detector in normal WIMP search operations to reduce the rate of rock neutrons \cite{Battat2017} reaching the fiducial volume of the detector was removed, with the exception of the underfloor shielding which could not be removed for mechanical reasons.  The $^{252}$Cf source was place in a cylindrical lead canister to shield the detector against gammas produced in the fission process.

\section{Data analysis and track reconstruction}\label{sec:dataanalysis}
Nuclear recoil candidates were selected using the cuts described in a previous analysis of the same data \cite{Battat2016}, and the ionization energy calibration in number of ion pairs (NIPs) was obtained using 5.9 keV X-rays from an $^{55}$Fe source, a procedure also detailed in that work. Because of the quenching factor, the associated recoil energy depends on the nuclear species (carbon, fluorine and sulfur)~\cite{Hitachi2008}.

Once the nuclear recoils are selected, their three-dimensional information can be extracted. This track information has its $x$ component along the grid wires, $y$ components along the anode wires and $z$ component along the drift direction. These three range components are called $\Delta x$, $\Delta y$ and $\Delta z$, with the reconstructed 3D range given by R$_3 = \sqrt{\Delta x^2+\Delta y^2 + \Delta z^2}$. The methods used to reconstruct each range component are described below. 

\subsection{Reconstruction of track $\Delta x$ range}\label{sec:deltax}
A point-like ionization cloud (zero range) positioned symmetrically between two anode wires would generate signal on both wires. Likewise, an ionization track of 2\,mm extent in the $x$-direction could generate a signal on a single wire. To establish a single, consistent definition of $\Delta x$ for each recoil, we count the number of anode wires $n$ with signal above threshold, multiply by the 2 mm wire pitch, and subtract one-half of the pitch \cite{Burgos2009a}:
\begin{equation}
\Delta x =  \left( 2n - 1 \right) \si{\milli\meter} \,.
\label{eq:delta.x} 
\end{equation} 
\nolinebreak

\subsection{Reconstruction of track $\Delta y$ range}\label{sec:deltay}
After an interaction inside the detector, cations resulting from signal avalanches near the anode wires drift to the grid wires and induce voltage signals. In this process, the highest voltage signal pulse occurs on the grid wire that is nearest to the position of the avalanche. The induced signal charge pulses on other contiguous grid wires which are further away from the position of the avalanche grow progressively smaller \cite{Blum2008}.  Hence, the $\Delta y$ range can be extracted by determining how the induced charge is shared between the grid wires as a function of time. 

\begin{figure} [b!] 
\centering
\includegraphics[width=.5\textwidth,height=0.33\textheight]{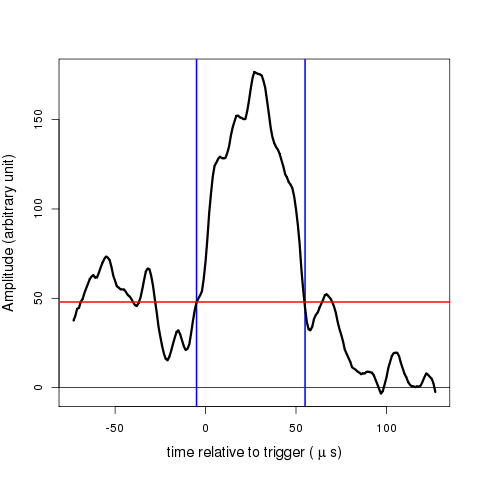} 
\caption{Integral grid waveform for a double hit event. The two blue lines are the start $t_s$ and end $t_e$ times of the region of interest defined by 25$\%$ (red line) of the maximum pulse amplitude of the resultant main ionisation charge cloud.}
\label{fig:deltaypulse}
\end{figure}

To determine the $\Delta y$ range component for each of the events that passed all the analysis cuts, the profile of the sum of charge from all the grid channels that recorded ionization hits (integral grid waveform) was used, as illustrated in Figure \ref{fig:deltaypulse}.  This is using the $y$ information at the start $t_s$ and end $t_e$ times of the integral grid waveform defined by the region of interest (25$\%$ of the maximum pulse amplitude) of the main event charge cloud. 
Signals over the 25$\%$ mark of the maximum pulse amplitude was used to reduce the effect of noise on the $\Delta y$ measurements. The $y$ component of the main charge cloud at each of these times $y_t$ was deduced using: 
\begin{equation}
y_{t} =  \left( \frac{\sum\limits_{j=1}^{n} 2jV_{t_j} }{V_{t} } \right) \si{\milli\meter} \,,
\label{eq:yt} 
\end{equation} 
\nolinebreak
where $V_{t_j}$ is the voltage signal deposited on $j^{th}$ grid wire at time $t$ for an event with a total number of $n$ grid wire hits. The $V_{t}$ parameter is the total voltage signal recorded on all the $n$ grid wires by an event at a given time, while the factor 2 is due to the grid wire pitch of 2~\si{\milli\meter}. Then, an event $\Delta y$ parameter can be estimated as:
 \begin{equation}
\Delta y =  | y_{t_s} - y_{t_e} | \,.
\label{eq:deltay} 
\end{equation} 
\nolinebreak
The $y_{t_s}$ and $y_{t_e}$ are the $y_t$ informations recorded at times $t_s$ and $t_e$, respectively. The expectation is that the maximum and minimum $y_t$ values should be recorded at $t_s$ and $t_e$ times.
 
\subsection{Reconstruction of track $\Delta z$ range}\label{sec:deltaz}
To reconstruct the $\Delta z$ range component, the integral anode waveform is first generated by summing together the waveforms from all anode channels that registered ionization from a given recoil. Next, $\Delta z$ is computed from:
\begin{equation}
\Delta z = \left(v_{I} \times \mbox{FWHM} \right) \si{\milli\meter},
\label{eq:deltaz}
\end{equation}
\begin{figure} [b!]  
\centering
\includegraphics[width=.5\textwidth,height=0.33\textheight]{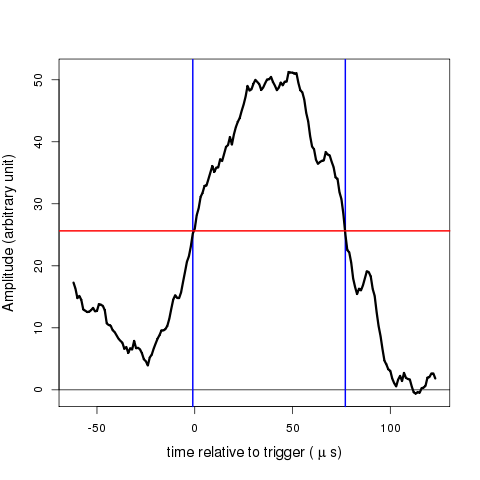} 
\caption{Integral anode waveform for a multiple hit event, showing the analysed main charge cloud and the determined full width half maximum (FWHM) parameter. The red line marks 50\si{\percent} of charge amplitude in the main charge cloud, the two blue lines are the lower and upper temporal bounds of the FWHM.}
\label{fig:fwhmipeak}
\end{figure}
where FWHM is the full-width-half-maximum of the main ionization peak in the integral waveform (the $I$-peak, as defined in Ref.~\cite{Battat2016, Snowdenifft2014}), and $v_{I}$ is the drift velocity of those charge-carriers (see Figure \ref{fig:fwhmipeak}).
The 50\% charge threshold of the resultant waveform was chosen similar to Ref. \cite{Burgos2009a}.
 
\subsection{Optimal directions and the oscillation parameter}\label{sec:otherparam}
The intrinsic strength of the directional axial range can be analysed using the $\delta$ parameters defined in Equation \ref{eq:axialcompdiffx} below, as the source was moved from the optimal to the anti-optimal directions.  To understand how these axial range component parameters vary from the average values for events in the optimal and the anti-optimal directions, $\delta x_{op}$, $\delta y_{op}$, $\delta z_{op}$ and $\delta x_{ao}$, $\delta y_{ao}$, $\delta z_{ao}$ values were computed, respectively. This was done by subtracting the average range component results obtained from the anti-optimal exposures from result of each of the runs.  For instance, the anti-optimal direction for the $\Delta x$ range are the $\widehat{y}$ and $\widehat{z}$ directions. Similarly, the anti-optimal directions for the $\Delta y$ ($\Delta z$) are $\widehat{x}$ and $\widehat{z}$ ($\widehat{x}$ and $\widehat{y}$) directions. Hence the $\delta x_{op}$ and $\delta x_{ao}$ parameters can be defined as:
\begin{align}
\label{eq:axialcompdiffx} 
\begin{split}
\delta x_{op} &= \left< \Delta x\right>_{x} - \frac{1}{2}\left( \left<\Delta x\right>_y + \left<\Delta x\right>_z \right),
\\ 
\delta x_{ao} &= \frac{1}{2}\left|  \left<\Delta x\right>_y - \left<\Delta x\right>_z    \right|, 
\end{split}
\end{align}
where $\delta x_{op}$ quantifies the sensitivity of DRIFT-IId to recoil tracks whose mean direction is parallel to the $\widehat{x}$ (optimal axis for $\Delta x$), while $\delta x_{ao}$ is a control statistic for the anti-optimal axes whose values should be consistent with zero.

A powerful statistic to search for the sidereal oscillation in the WIMP wind direction is the ratio of the range component $\Delta z / \Delta x$ along the $\widehat x$ and $\widehat z$ directions \cite{Burgos2009a}.  For a given neutron exposure that produces $N$ recoil events, we define the average range component ratio as:
\begin{equation}
\left< \frac{\Delta z}{\Delta x} \right> = \frac{1}{N} \sum_{i=1}^{N} \frac{\Delta z_i}{\Delta x_i}.
\label{eq:deltazdeltaxoverline}
\end{equation}
We then define an oscillation amplitude $A$ as the difference in this ratio for neutron exposures oriented along $\widehat{z}$ and $\widehat{x}$ divided by the mean of the ratios:
\begin{equation}
A = \frac{ \left<\frac{\Delta z}{\Delta x}\right>_{z} - \left<\frac{\Delta z}{\Delta x}\right>_{x}  }{ \frac{1}{2}\left(  \left<\frac{\Delta z}{\Delta x}\right>_{z} + \left<\frac{\Delta z}{\Delta x}\right>_{x} \right)}.
\label{eq:magoscillation}
\end{equation}
The $A$ parameter quantifies the amplitude of the modulation in the directional signal that DRIFT would see over the 12 sidereal hour period shown in Figure \ref{fig:directionalsignal}.

\section{Results and Discusions}\label{sec:axialresult}
\begin{figure}[t] 
\centering
\subfigure[$\Delta x$ from $-x$ run.]{
\includegraphics[width=0.3\linewidth,height=0.235\textheight]{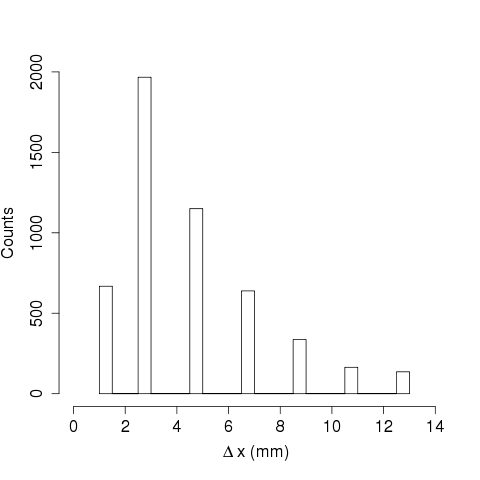} 
\label{fig:deltaxforx}
} \hfil
\subfigure[$\Delta x$ from $-y$ run.]{
\includegraphics[width=0.3\linewidth,height=0.235\textheight]{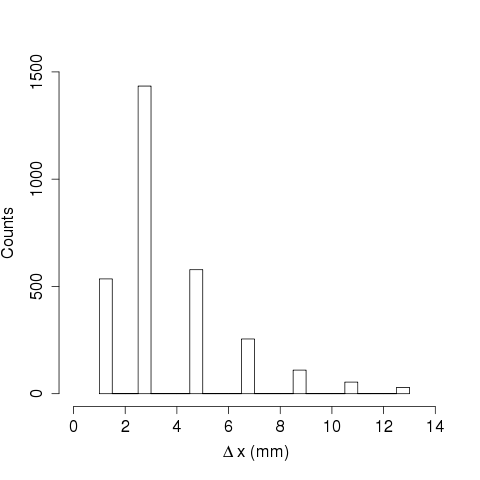} 
\label{fig:deltaxfory}
}
\subfigure[$\Delta x$ from $-z$ run.]{
\includegraphics[width=0.3\linewidth,height=0.235\textheight]{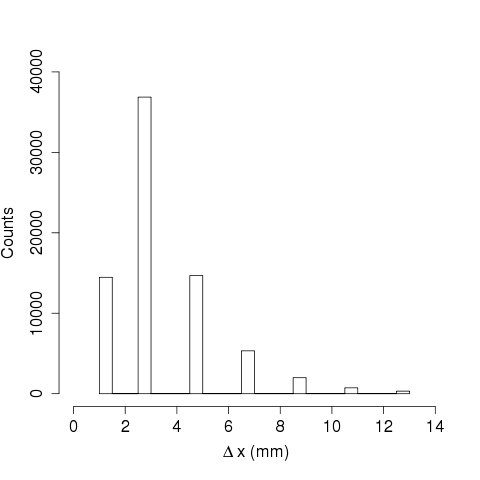} 
\label{fig:deltaxminusz}
}
\subfigure[$\Delta y$ from $-x$ run.]{
\includegraphics[width=0.3\linewidth,height=0.235\textheight]{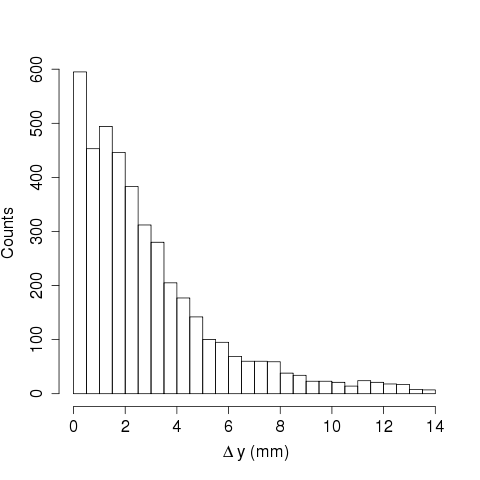} 
\label{fig:deltayforx}
} \hfil
\subfigure[$\Delta y$ from $-y$ run.]{
\includegraphics[width=0.3\linewidth,height=0.235\textheight]{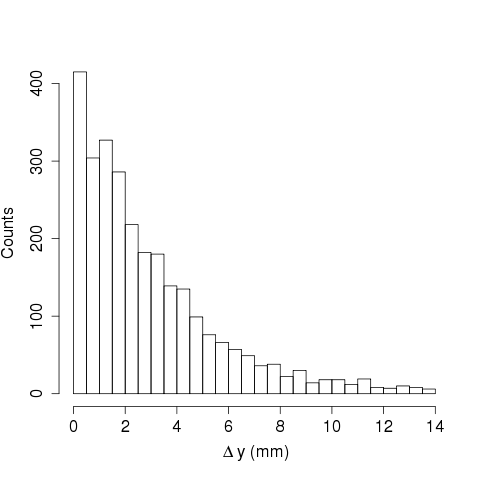} 
\label{fig:deltayfory}
}
\subfigure[$\Delta y$ from $-z$ run.]{
\includegraphics[width=0.3\linewidth,height=0.235\textheight]{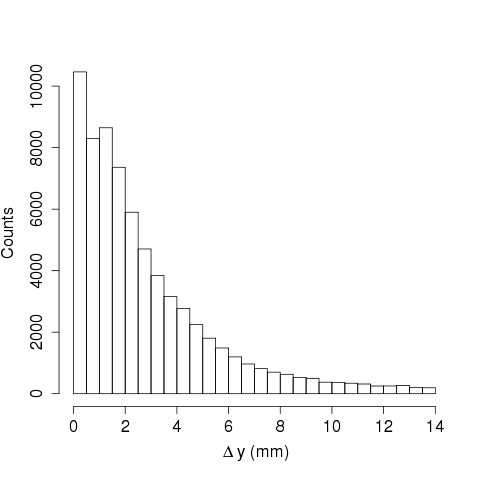} 
\label{fig:deltayminusz}
}
\subfigure[$\Delta z$ from $-x$ run.]{
\includegraphics[width=0.3\linewidth,height=0.235\textheight]{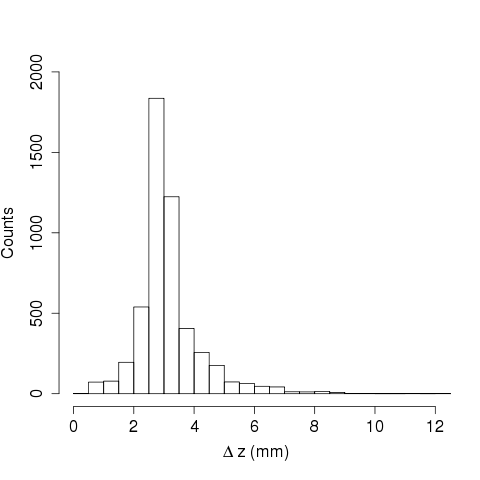} 
\label{fig:deltazforx}
} \hfil
\subfigure[$\Delta z$ from $-y$ run.]{
\includegraphics[width=0.3\linewidth,height=0.235\textheight]{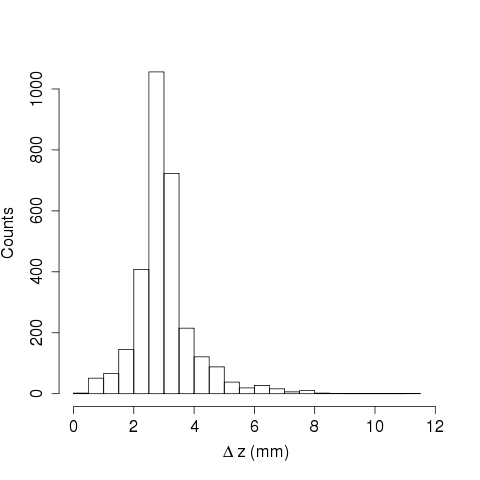} 
\label{fig:deltazfory}
}
\subfigure[$\Delta z$ from $-z$ run.]{
\includegraphics[width=0.3\linewidth,height=0.235\textheight]{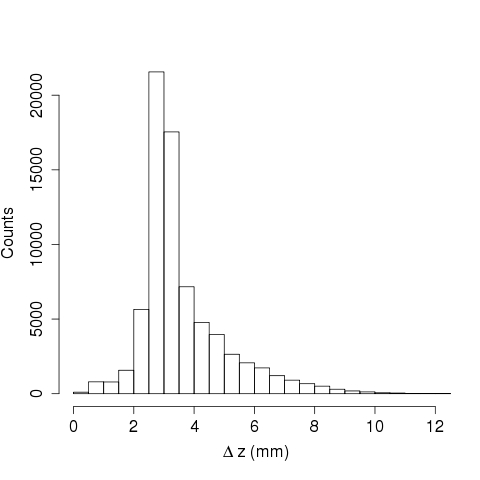} 
\label{fig:deltazminusz}
}
\caption{Distributions of $\Delta x$, $\Delta y$ and $\Delta z$ nuclear recoil axial range components obtained from different directions of neutron exposure. The top panels show the $\Delta x$ parameters, the middle panels are the $\Delta y$ parameters while the bottom panels are the $\Delta z$ parameters obtained from $-x$, $-y$, and $-z$ neutron exposures, respectively.}
\label{fig:distforalldeltaparams}
\end{figure}

The distribution of measured range components from each of the three neutron exposures is shown in Figure \ref{fig:distforalldeltaparams}, with the means and standard errors on the means reported in Table \ref{table:axialparameter}.
\begin{table}[t] 
\small
\centering
\caption{Axial range components of nuclear recoil tracks obtained from $-x$, $-y$, and $-z$ directed neutron runs. The mean neutron directions (MND) are shown in the first column, number of recoil candidates that pass analysis cuts $N$ are in the second column while run live-time for each of the neutron exposures are in the third column. In columns four to six are the average $\Delta x$, $\Delta y$, and $\Delta z$ parameters for each of run. The $\left<\frac{\Delta z}{\Delta x}\right>$ parameters obtained for each of these runs is shown in column seven. Quoted errors are 1$\sigma$ statistical uncertainties. Event energies are in the range of about 500$\leq$NIPs$\leq$6000.} 
\begin{adjustbox}{width= 15.2 cm}
\tiny
\begin{tabular}{l*{7}{c}r}
\hline \hline \\
MND    	& 	     $N$  	&    live-time (days)   &  	$\left<\Delta x\right>$~(\si{\milli\meter})  & $\left<\Delta y\right>$~(\si{\milli\meter}) 	& 	 $\left<\Delta z\right>$ ~(\si{\milli\meter}) 	 &        $\left<\frac{\Delta z}{\Delta x}\right>$    	 \\ \\
\hline 
\\
$-x$       	&  	    5074 	&	   0.915             &    	 4.634$\pm$0.044		&    2.626$\pm$0.043        	&       	3.261$\pm$0.016			&	0.965$\pm$0.010			\\ \\
	
$-y$      	&    	    2857 	&	   0.561	        &   	 3.885$\pm$0.044		 &   2.827$\pm$0.055	        & 		3.002$\pm$0.018			&	1.092$\pm$0.014		  	 \\ \\ 

$-z$      	&    	    69960	&	    3.713	        &   	 3.716$\pm$0.008     	&    2.569$\pm$0.012      		&    	  	3.586$\pm$0.005  	 		&	1.331$\pm$0.004			 \\ \\ 

\hline \hline
\end{tabular} 
\end{adjustbox}
\label{table:axialparameter}
\vspace*{0.25 cm}
\end{table} 
In general, it can be seen in this table that the values of the $\Delta x$ parameter are bigger for each of the runs compared to their corresponding $\Delta y$ and $\Delta z$ range parameters regardless of orientation. This is because of the different methods and systematics used in each of the reconstruction as discussed in Section \ref{sec:dataanalysis}. 

It can be seen that all the $\Delta x$, $\Delta y$ and $\Delta z$ results obtained from this new data set are larger than the results of a similar study of sulfur recoils in pure CS$_2$ \cite{Burgos2009a}. In fact, the average R$_3$ parameter obtained from $x$, $y$ and $z$ exposures in this measurement is larger than results obtained with only sulfur tracks in Ref. \cite{Burgos2009a} by a factor of 1.7$\pm$0.2, 1.6$\pm$0.2 and 1.5$\pm$0.1, respectively. This is similar to the average R$_3$ parameter from fluorine and carbon to sulfur recoil track range  ratios of 1.2 to 1.4, predicted by SRIM \cite{Ziegler2010} for events of 500 to 6000 NIPs.  It can be seen that each of the axial range component parameters shown in Table \ref{table:axialparameter} returned a larger value where the MND is oriented in a direction parallel to the respective axes of the range components.  For instance, the $\Delta x$ parameter obtained from $-x$ directed neutron exposure is greater than the results obtained from $-z$, and $-y$ exposures.  This is because the MND in this direction is perpendicular to the orientation of the anode wires. This is the optimal direction of the detector for $\Delta x$ measurements while the other two perpendicular directions correspond to the anti-optimal directions for the $\Delta x$ measurements.  The same is true for the other range components $\Delta y$ and $\Delta z$.

\begin{figure} [t] 
\centering
\includegraphics[width=.55\textwidth,height=0.385\textheight]{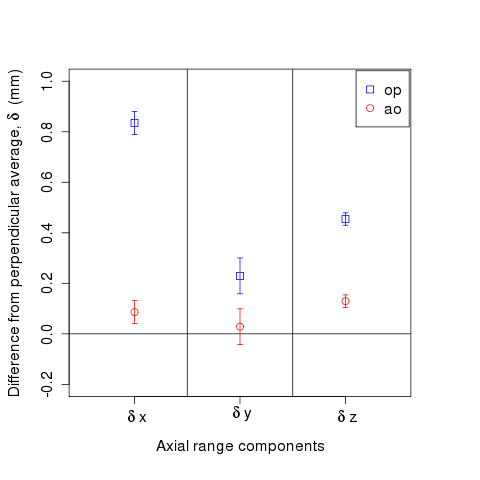} 
\caption{The $\delta x$, $\delta y$ and $\delta z$ parameters obtained from $-x$, $-y$ and $-z$ exposures. Results from optimal directions are shown with blue square points while the red circular points are results from the anti-optimal directions.  Error bars are 1$\sigma$ statistical uncertainties.}
\label{fig:axialcompdiff}
\end{figure}

The difference between these axial range components in a particular direction and average results from events in the two perpendicular anti-optimal directions (detector axes), the $\delta x$, $\delta y$ and $\delta z$ parameters, were computed as defined in Equation \ref{eq:axialcompdiffx}. The results from these analyses are depicted in Figure \ref{fig:axialcompdiff}.
\begin{figure}[t!] 
\centering
\includegraphics[width=0.55\linewidth,height=0.385\textheight]{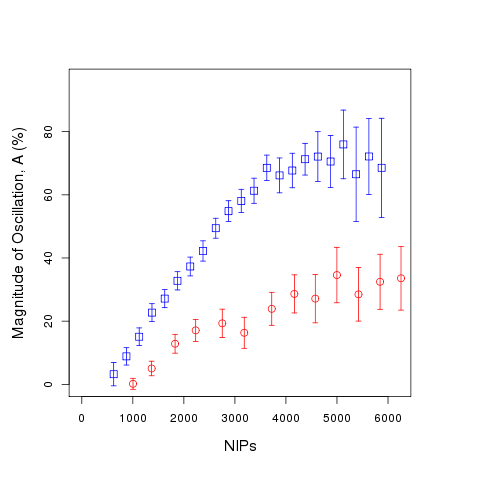} 
\caption{Magnitude of the oscillation amplitude $A$ for $-x$ and $-z$ directed neutron runs as a function of NIPs with 250-NIPs bins. For comparison, the results of a similar study for sulfur recoils in pure CS$_2$ is shown as red circles.}
\label{fig:magofoscillation}
\end{figure}
As expected, it was found that the maximum $\delta x$, $\delta y$ and $\delta z$ parameters were observed from the $\Delta x$, $\Delta y$ and $\Delta z$ axial range components in the $-x$, $-y$ and $-z$ directed neutron exposures, respectively.  

Using Equation \ref{eq:deltazdeltaxoverline}, the magnitude of oscillation present as the source is swapped between $\widehat{z}$ and $\widehat{x}$ directions was investigated. Due to the low signal of the $\Delta y$ range, we omit it in this and further axial directional computations.  The $\left<\frac{\Delta z}{\Delta x}\right>$ results are shown for each of the directions of exposures in column seven of Table \ref{table:axialparameter}. It can be seen, that the maximum $\left<\frac{\Delta z}{\Delta x}\right>$ values were observed from events in the $z$ exposures and minimum results from $-x$ events as expected, due to larger $\Delta x$ parameters observed in this axis.  Using these measurements, a peak-to-peak oscillation magnitude of 0.366$\pm$0.011 can be seen by comparing the $\left<\frac{\Delta z}{\Delta x}\right>$ results from $-z$ and $-x$ exposure directions. This results in an average oscillation of 34$\sigma$ as the source is interchanged between these two neutron source locations for the 75,034 neutron events.  

The oscillation amplitude $A$ was computed as a function of NIPs, with bin size of 250 NIPs. The results are shown in Figure \ref{fig:magofoscillation} for recoils from $-x$ and $-z$ directed neutron runs. For NIPs conversion to carbon, fluorine and sulfur equivalent energies, see Table 1 in Ref. \cite{Battat2017}. As expected, it can be seen that the magnitude of the  $A$ oscillation obtained by moving the source from $-x$ to $-z$ neutron exposure directions increases with the event's recoil ionization. 

Compared with our previous study of the oscillation amplitude for sulfur recoil tracks in pure CS$_2$ \cite{Burgos2009a}, we find that with the CS$_2$:CF$_4$:O$_2$ gas mixture, $A$ is, on average, a factor of 2.4 larger. These new results are consistent with the expected increase in the average R$_3$ range and consequently $\left<\frac{\Delta z}{\Delta x}\right>_z - \left<\frac{\Delta z}{\Delta x}\right>_x$ due to the inclusion of longer nuclear recoil tracks from the carbon and fluorine components of the target gas in this analysis. 

\section{Conclusion}
The sensitivity of the DRIFT-IId detector to the axial range components of low-energy nuclear recoils was investigated.  Using a $^{252}$Cf neutron source with mean neutron direction aligned with the $x$, $y$ and $z$ detector axes, and the current DRIFT-IId WIMP-search gas mixture of CS$_2$:CF$_4$:O$_2$, we demonstrated for the first time that the range component signature could be reconstructed. The prevalence of carbon and fluorine recoils in the data meant that the average measured recoil range was 50\% larger than in the case of sulfur recoils in pure CS$_2$. These longer tracks enable a more accurate reconstruction, leading to an improved sensitivity to the range oscillation amplitude $A$. These measurements demonstrate that the addition of oxygen to the target gas mixture has not degraded DRIFT's directional sensitivity. For dark matter search operations, these measurements can be used to track the mean directions of potential positive signals that may reach the detector from the direction of the Cygnus constellation over a sidereal day.

\acknowledgments
The DRIFT collaboration is very grateful to the NSF and Cleveland Potash Ltd for their funding and continued support, respectively. We acknowledge support from the STFC through grant no. ST/P00573X/1. JBRB acknowledges the support of the National Science Foundation (EAGER PHY-1649966), the Research Corporation Cottrell College Science Award (\# 23325) and the Sloan Research Fellowship (BR2012-011). DL acknowledges the support of the National Science Foundation (Grant Nos. 1103420 and 1407773). DPSI is grateful to the National Science Foundation for support through grant no. 1506237.

\end{document}